\documentclass[letterpaper,11pt]{article}
\usepackage{amsmath}
\usepackage[dvips]{graphicx}
\begin{document}
\title{Newtonian and General Relativistic Models of Spherical Shells}
\author{D. Vogt\thanks{e-mail: dvogt@ime.unicamp.br} 
\and
P. S. Letelier\thanks{e-mail: letelier@ime.unicamp.br}\\
Departamento de Matem\'{a}tica Aplicada-IMECC, Universidade \\ 
Estadual de Campinas 13083-970 Campinas, S.\ P., Brazil}
\maketitle
\begin{abstract}
A family of spherical shells with varying thickness is derived by using a simple Newtonian 
potential-density pair. Then, a particular isotropic form of a metric in spherical coordinates is used 
to construct a General Relativistic version of the Newtonian family of shells. The matter of these relativistic 
shells presents equal azimuthal and polar pressures, while the radial pressure is a constant times  
the tangential pressure. We also make a first study of stability of both the Newtonian and relativistic families 
of shells. 

\end{abstract}
\section{Introduction}

Distributions of matter in form of shells have been a useful tool in astrophysics and General
Relativity, often providing simplified but analytically tractable models in cosmology,
gravitational collapse and supernovae. In General Relativity, the pioneering work on 
constructing solutions of the Einstein field equations representing shells was done
by Israel \cite{i66}. In his formalism, the shells are constructed by joining two manifolds 
across a three-dimensional time-like surface, and the matter representing the shell 
is interpreted as a discontinuity of the extrinsic curvature on the surface. This formalism 
has been used to study several models of relativistic shells.   
\cite{k06} give a brief review of some of these models and also discuss shells which admit 
closed equations of state, including non-linear polytropic models.  

An exact solution of a static shell surrounding a black hole was obtained by \cite{fhk90}, 
and the stability of such a configuration against radial perturbations was studied by \cite{blp91},  
while non-radial perturbations were analysed by \cite{s98}. Non-radial perturbations of self-gravitating 
static fluid shells in the Newtonian gravity were studied by \cite{bs99}.  

In this work, we propose a Newtonian analytical potential-density pair that represents
a family of spherical shells with varying thickness. For this potential, we calculate the rotation curves
of test particles and make a first study of stability of the shells by considering the stability of
circular orbits of the test particles. This is done in Section \ref{sec_ns}. In Section \ref{sec_gr}, we
consider a General Relativistic version of the Newtonian shells. For this task, we employ 
a particular isotropic form of a metric in spherical coordinates. We calculate the components 
of the energy-momentum tensor of the shell's matter and also analyse the geodesic circular motion 
of test particles and discuss their stability. Finally, in Section \ref{sec_dis} we summarize our 
results.    
\section{A Model of a Newtonian Shell} \label{sec_ns}

\begin{figure}
\centering
\includegraphics[scale=0.72]{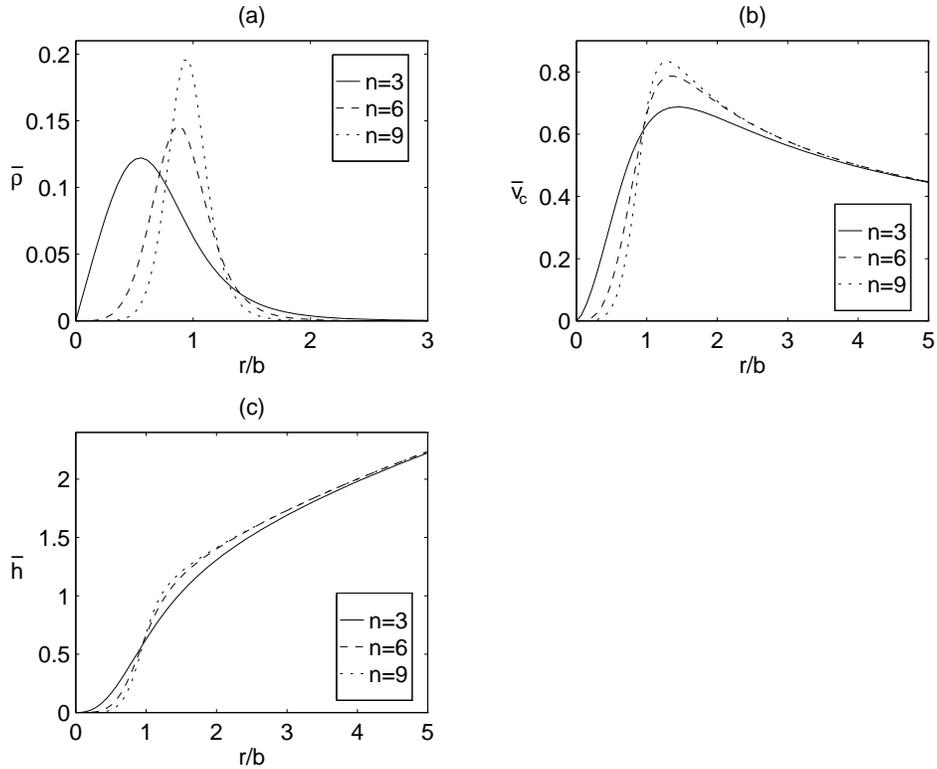}
\caption{(a) The mass density $\bar{\rho}=\rho/(m/b^3)$ (equation \ref{eq_rho_n}), 
as function of $r/b$ for $n=3$, $n=6$ and $n=9$. (b) The circular velocity 
$\bar{v}_c=v_c/(Gm/b)^{1/2}$ (equation \ref{eq_vc_n}), and (c) the angular
momentum $\bar{h}=h/(Gmb)^{1/2}$ (equation \ref{eq_h_n})  as functions of 
$r/b$ for $n=3$, $n=6$ and $n=9$.} \label{fig1}
\end{figure}

In this section, we present a simple potential-density pair that describes
a family of spherical shells of matter. We shall consider the following potential:
\begin{equation} \label{eq_phi_n}
\Phi=-\frac{Gm}{\left( r^n+b^n \right)^{1/n}} \mbox{,}
\end{equation}
where $b>0$ is a parameter with dimensions of length and $n$ is a non-negative integer.
By using the Poisson equation in spherical coordinates,
\begin{equation} \label{eq_poisson}
\frac{1}{r^2}\frac{\mathrm{d}}{\mathrm{d}r} \left( 
r^2 \frac{\mathrm{d}\Phi}{\mathrm{d}r} \right) =4\pi G \rho \mbox{,}
\end{equation}
one obtains a mass density given by
\begin{equation} \label{eq_rho_n}
\rho=\frac{m(n+1)b^nr^{n-2}}{4\pi \left( r^n+b^n \right)^{2+1/n}} \mbox{.}
\end{equation}
For $n=2$, the pair (\ref{eq_phi_n}) and (\ref{eq_rho_n}) is reduced to the 
Plummer model \cite{pl11,bt08} that has a monotone decreasing density
profile. However, for $n>2$ the density vanishes on $r=0$ and thus we 
have a shell-like distribution of matter. Fig.\ \ref{fig1}(a) shows some curves of
the dimensionless mass density $\bar{\rho}=\rho/(m/b^3)$ as function of 
$r/b$ for $n=3$, $n=6$ and $n=9$. For higher values of $n$, the mass 
density becomes more concentrated (thinner shell), with maximum at
\begin{equation}
r_{max}=\left( \frac{n-2}{n+3} \right)^{1/n}b \mbox{.} 
\end{equation} 
For large values of $r$, the density (\ref{eq_phi_n}) decays as $1/r^{n+3}$, 
thus very fast for large $n$, as can be seen in Fig.\ \ref{fig1}(a).  

The circular velocity $v_c=(r\Phi_{,r})^{1/2}$ and angular momentum per unit 
mass $h=rv_c$ of test particles moving in circular orbits are given by 
\begin{gather} 
v_c=\sqrt{\frac{Gmr^n}{\left( r^n+b^n \right)^{1+1/n}}} \label{eq_vc_n} \mbox{,} \\
h=\sqrt{\frac{Gmr^{n+2}}{\left( r^n+b^n \right)^{1+1/n}}} \label{eq_h_n} \mbox{.} 
\end{gather}
Curves of the circular velocity $\bar{v}_c=v_c/(Gm/b)^{1/2}$ and of the angular
momentum $\bar{h}=h/(Gmb)^{1/2}$ for $n=3$, $n=6$ and $n=9$ are depicted in 
Figs  \ref{fig1}(b) and (c), respectively.
The stability of circular orbits can be determined using the Rayleigh criterion
of stability for a rotating fluid (\cite{r17}; see also \cite{ll87}), and
that can be adapted to study
circular orbits of test particles in self-gravitating systems (see e.\ g.\  
\cite{let03} for an application of the Rayleigh criterion of stability in 
General Relativity). The condition for a stable orbit is
\begin{equation}
h\frac{\mathrm{d}h}{\mathrm{d}r} >0 \mbox{.}
\end{equation}
Using equation (\ref{eq_h_n}), it is straigthforward to check that the orbits 
are always stable.

It is instructive to consider the limit $n \rightarrow \infty$. In this case, 
we have an infinitesimal thin shell on $r=b$, and the limits of the 
potential (\ref{eq_phi_n}) and circular velocity (\ref{eq_vc_n}) are
\begin{gather}
\Phi=\begin{cases} 
-\frac{Gm}{b}& \text{if  } 0 \leq r<b, \\
-\frac{Gm}{r}& \text{if  } r \geq b, 
\end{cases} \\
v_c=\begin{cases} 
0 & \text{if  } 0 \leq r<b, \\
\sqrt{\frac{Gm}{r}} &\text{if  } r \geq b.
\end{cases}
\end{gather}
Thus, for $r<b$ the potential is constant and a test particle is at rest 
inside the shell, and if $r \geq b$ the shell exerts on the test particle the
same gravitational force as a point mass located on $r=0$, in agreement with
Newton's Shell Theorem.
\section{A Model of a General Relativistic Shell} \label{sec_gr}

Now, we study a General Relativistic version of the Newtonian potential-density
pair discussed in the previous section. For this, it is convenient to choose a particular metric
in isotropic form in spherical coordinates $(t,r,\theta,\varphi)$, which we write as
\begin{equation} \label{eq_metric}
\mathrm{d}s^2=\left( \frac{1-f}{1+f} \right)^2c^2\mathrm{d}t^2-\left( 1+f \right)^4 
\left( \mathrm{d}r^2+r^2\mathrm{d}\theta^2
+r^2\sin^2 \theta \mathrm{d} \varphi^2 \right) \mbox{,}
\end{equation}
where $f=f(r)$. In this particular form of metric, the Schwarzschild solution
is given by $f=Gm/(2c^2r)$. A similar form of the
metric (\ref{eq_metric}) in cylindrical coordinates was used by \cite{vl05} 
to construct General Relativistic versions of Newtonian models for distributions of 
matter in galaxies.

For metric (\ref{eq_metric}), the Einstein equations $G_{\mu\nu}=-(8\pi G/c^4)T_{\mu\nu}$
yield the following expressions for the non-zero components of the energy-momentum 
tensor $T_{\mu\nu}$:
\begin{gather}
T^t_t=-\frac{c^4}{2\pi G \left( 1+f \right)^5} \frac{1}{r^2} 
\frac{\mathrm{d}}{\mathrm{d}r} \left( r^2\frac{\mathrm{d}f}{\mathrm{d}r} \right) \mbox{,} \label{eq_Ttt} \\
T^r_r=\frac{c^4}{2\pi G \left( 1+f \right)^5\left( 1-f \right)} \frac{\mathrm{d}f}{\mathrm{d}r} 
\left( \frac{f}{r}+\frac{\mathrm{d}f}{\mathrm{d}r} \right) \mbox{,} \\
T^{\theta}_{\theta}=T^{\varphi}_{\varphi}=\frac{c^4}{4\pi G \left( 1+f \right)^5\left( 1-f \right)} 
\left[ f\frac{\mathrm{d}^2f}{\mathrm{d}r^2}+\frac{f}{r}\frac{\mathrm{d}f}{\mathrm{d}r}
-\left( \frac{\mathrm{d}f}{\mathrm{d}r} \right)^2 \right] \mbox{.} \label{eq_Tthth}
\end{gather}
Because $T_{\mu\nu}$ has only diagonal non-zero components, the energy density is 
directly given by $\epsilon=T^t_t/c^2$ and the pressures or tensions
along a direction $k$ read $P_k=-T^k_k$. The `effective Newtonian density'
$\rho_N=\epsilon+P_r/c^2+P_{\theta}/c^2+P_{\varphi}/c^2$ reads 
\begin{equation} \label{eq_rho_N1}
\rho_N=-\frac{c^2}{2\pi G \left( 1+f \right)^5\left( 1-f \right)} \frac{1}{r^2}
\frac{\mathrm{d}}{\mathrm{d}r} \left( r^2\frac{\mathrm{d}f}{\mathrm{d}r} \right) \mbox{.}
\end{equation}
The relation between the function $f(r)$ and the Newtonian potential $\Phi(r)$ 
is obtained by comparing equation (\ref{eq_rho_N1}) or (\ref{eq_Ttt}) with 
equation (\ref{eq_poisson}) in the non-relativistic limit when $f \ll 1$, 
\begin{equation} \label{eq_f_phi}
f=-\frac{\Phi}{2c^2} \mbox{.} 
\end{equation}

Now, using equation (\ref{eq_f_phi}) with the potential (\ref{eq_phi_n}), 
equations (\ref{eq_Ttt})--(\ref{eq_rho_N1}) can be cast as
\begin{gather} 
\bar{\epsilon}=\frac{(n+1)\bar{b}^n\bar{r}^{n-2}}{4\pi \xi^{2-4/n}
\left( 1+\xi^{1/n} \right)^5} \mbox{,} \label{eq_eps} \\
\bar{P}_r=\frac{\bar{b}^n\bar{r}^{n-2}}{4\pi \xi^{2-4/n}
\left( -1+\xi^{1/n} \right) \left( 1+\xi^{1/n} \right)^5} \mbox{,} \label{eq_P_r} \\
\bar{P}_{\theta}=\bar{P}_{\varphi}=\frac{n\bar{b}^n\bar{r}^{n-2}}{8\pi \xi^{2-4/n}
\left( -1+\xi^{1/n} \right) \left( 1+\xi^{1/n} \right)^5} \mbox{,} \label{eq_P_phi} \\
\bar{\rho}_N=\frac{(n+1)\bar{b}^n\bar{r}^{n-2}}{4\pi \xi^{2-5/n}
\left( -1+\xi^{1/n} \right) \left( 1+\xi^{1/n} \right)^5} \mbox{,} \label{eq_rho_N2}
\end{gather}
where the dimensionless variables and parameters are as follows: $\bar{r}=r/r_s$, $\bar{b}=b/r_s$, 
$\bar{\epsilon}=\epsilon/(m/r_s^3)$, $\bar{\rho}_N=\rho_N/(m/r_s^3)$, $\bar{P}_k=P_k/(mc^2/r_s^3)$,
and where we defined $r_s=Gm/(2c^2)$ and $\xi=\bar{r}^n+\bar{b}^n$. We assume again that $n>2$. For $n=2$,  
equations (\ref{eq_eps})--(\ref{eq_rho_N2}) reduce to the perfect fluid sphere derived 
by \cite{b64}, which has as equation of state similar to the classical polytrope of index 5.
Note that the relation between $\bar{P}_r$ and $\bar{P}_{\theta}$ is independent of the 
radius and is simply $\bar{P}_r=2\bar{P}_{\theta}/n$.

In order to be physically meaningful, the components of the energy-momentum 
tensor should satisfy the energy conditions. The strong energy condition states that 
$\rho_N \geq 0$, whereas the weak energy condition imposes the condition $\epsilon \geq 0$. 
The dominant energy condition requires $|P_k/\epsilon| \leq c^2$. Equations (\ref{eq_eps})--(\ref{eq_rho_N2}) 
show that  the weak energy condition is always satisfied, and the strong condition holds if $\bar{b} \geq 1$. 
This condition also ensures that we have pressures everywhere. The inequality $|P_r/\epsilon| \leq c^2$
is satisfied when $\bar{b} \geq (n+2)/(n+1)$, and we have $|P_{\theta}/\epsilon| \leq c^2$ if  $\bar{b} \geq (3n+2)/(2n+2)$. 
Thus, there exists an interval of the parameter $\bar{b}$ where all energy conditions are satisfied.

\begin{figure}
\centering
\includegraphics[scale=0.72]{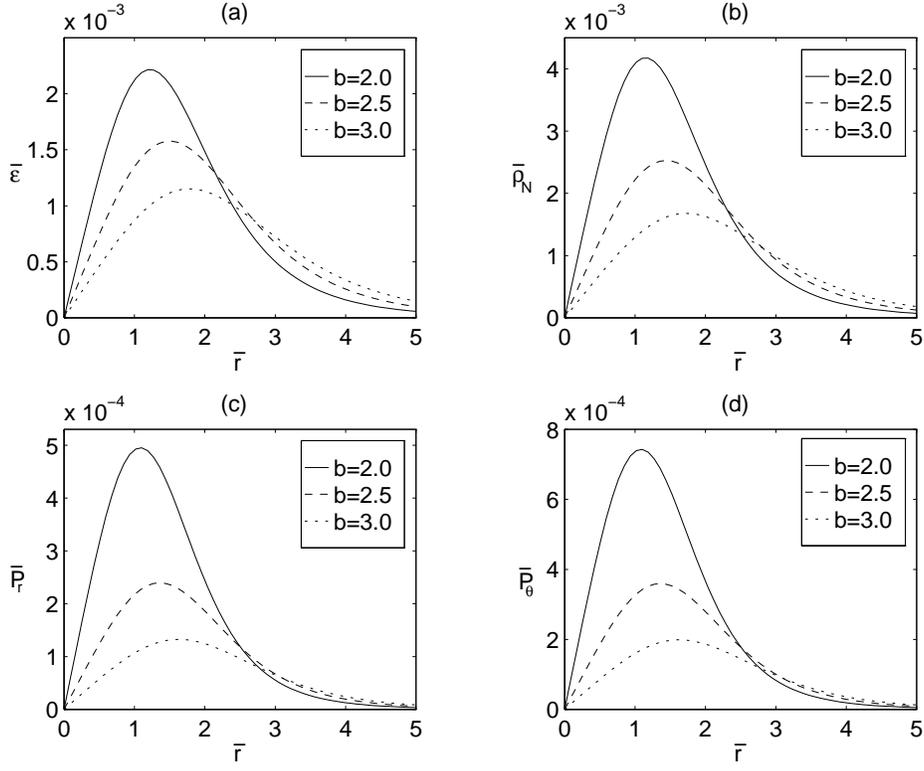}
\caption{(a) The energy density $\bar{\epsilon}=\epsilon/(m/r_s^3)$  (equation \ref{eq_eps}), 
(b) the effective Newtonian density $\bar{\rho}_N=\rho_N/(m/r_s^3)$  (equation \ref{eq_rho_N2}), 
(c) the radial pressure $\bar{P}_r=P_r/(mc^2/r_s^3)$  (equation \ref{eq_P_r}) and (d) the
azimuthal/polar pressure $\bar{P}_{\theta}=P_{\theta}/(mc^2/r_s^3)$  (equation \ref{eq_P_phi})
as functions of $\bar{r}=r/r_s$ for $n=3$, and parameter $\bar{b}=b/r_s=2$, $2.5$ and $3$.} \label{fig2}
\end{figure}

\begin{figure}
\centering
\includegraphics[scale=0.72]{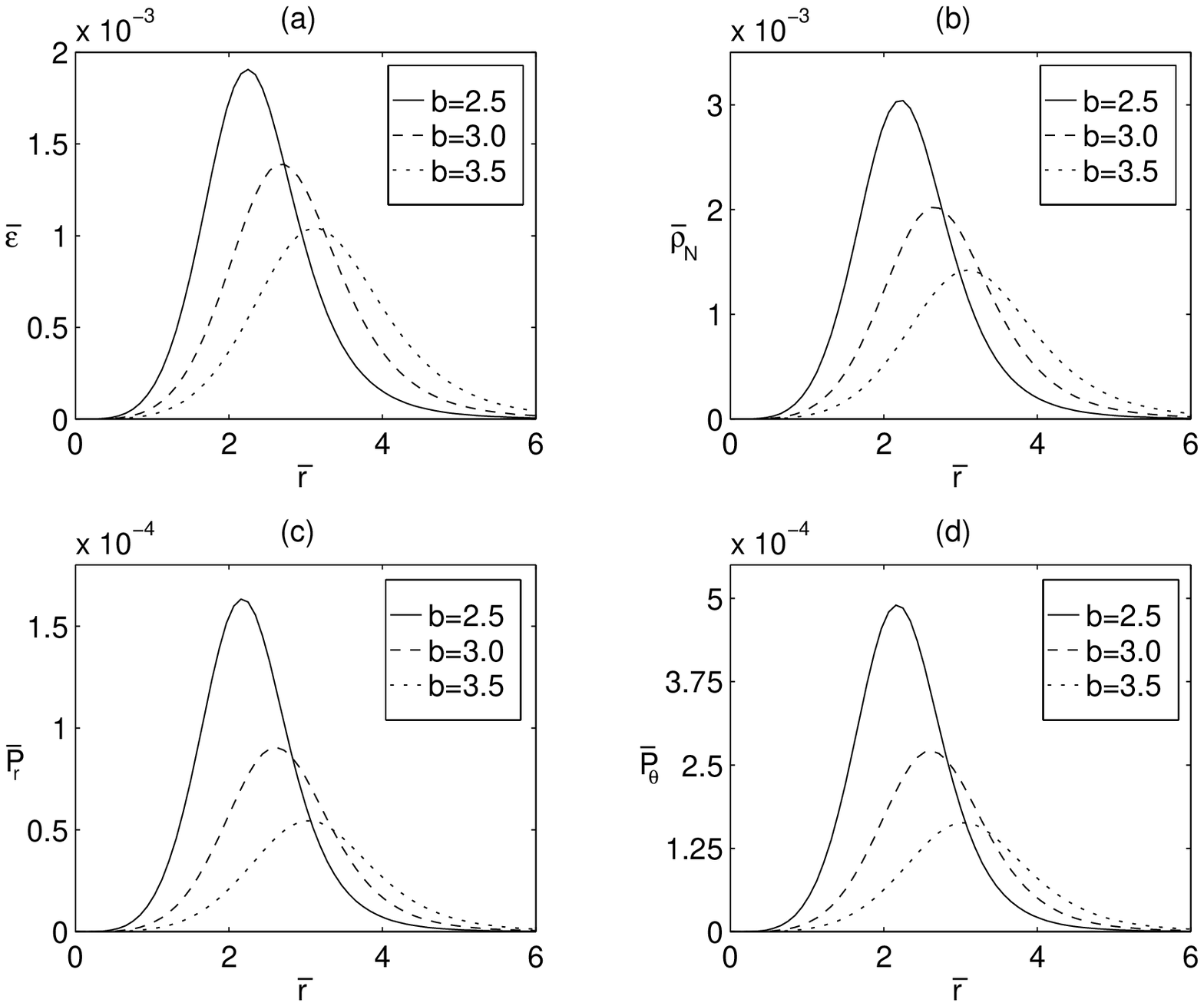}
\caption{(a) The energy density $\bar{\epsilon}=\epsilon/(m/r_s^3)$  (equation \ref{eq_eps}), 
(b) the effective Newtonian density $\bar{\rho}_N=\rho_N/(m/r_s^3)$  (equation \ref{eq_rho_N2}), 
(c) the radial pressure $\bar{P}_r=P_r/(mc^2/r_s^3)$  (equation \ref{eq_P_r}) and (d) the
azimuthal/polar pressure $\bar{P}_{\theta}=P_{\theta}/(mc^2/r_s^3)$  (equation \ref{eq_P_phi})
as functions of $\bar{r}=r/r_s$ for $n=6$, and parameter $\bar{b}=b/r_s=2.5$, $3$ and $3.5$.} \label{fig3}
\end{figure}

Figs \ref{fig2}(a)--(d) display, respectively, some curves of the energy density $\bar{\epsilon}$, the effective 
Newtonian density $\bar{\rho}_N$, the radial pressure $\bar{P}_r$ and the azimuthal/polar pressure
$\bar{P}_{\theta}$ as functions of the radial coordinate for the shell with $n=3$ and parameter values
$\bar{b}=2$, $2.5$ and $3$. In Figs \ref{fig3}(a)--(d), the same quantities are displayed for the 
shell with $n=6$ and parameter values $\bar{b}=2.5$, $3$ and $3.5$. All quantities have the profile of 
a shell-like distribution of matter. As the value of the parameter $\bar{b}$ is increased, the densities and 
pressures are smoothed out. The densities and pressures become more concentrated when $n$
is increased.  

As we did in the Newtonian model, we also study the circular orbits and stability of test particles. 
Without loss of generatity, if the geodesic motion is confined on the $\theta=\pi/2$ plane,
the expressions for the circular velocity and for the angular momentum are formally 
the same as in cylindrical coordinates, and we have \cite{vl05}
\begin{gather}
v_c=\sqrt{\frac{-2c^2rf_{,r}}{\left( 1-f \right) \left( 1+f+2rf_{,r} \right)}} \mbox{,} \label{eq_vc_r1} \\
h= cr^2 \left( 1+f \right)^2 \sqrt{\frac{-2f_{,r}}{r\left[ 1-f^2+2rf_{,r}\left( 2-f \right)\right] }} \mbox{.} \label{eq_h_r1}
\end{gather}
For the shell potential (\ref{eq_phi_n}), we get
\begin{gather}
\bar{v}_c=\sqrt{\frac{2\bar{r}^n\xi^{1/n}}
{\left( -1+\xi^{1/n} \right)
\left( \bar{b}^n-\bar{r}^n+\xi^{1+1/n} \right)}} \mbox{,} \label{eq_vc_r2} \\
\bar{h}=\frac{\sqrt{2}\bar{r}^{1+n/2}\left( 1+\xi^{1/n} \right)^2}
{\xi^{3/(2n)}\sqrt{\bar{r}^n-\bar{b}^n+\xi^{1+2/n}- 4\bar{r}^n\xi^{1/n}}} \mbox{,} \label{eq_h_r2} 
\end{gather} 
where $\bar{v}_c=v_c/c$, $\bar{h}=r_sc$ and $\xi=\bar{r}^n+\bar{b}^n$.

\begin{figure}
\centering
\includegraphics[scale=0.72]{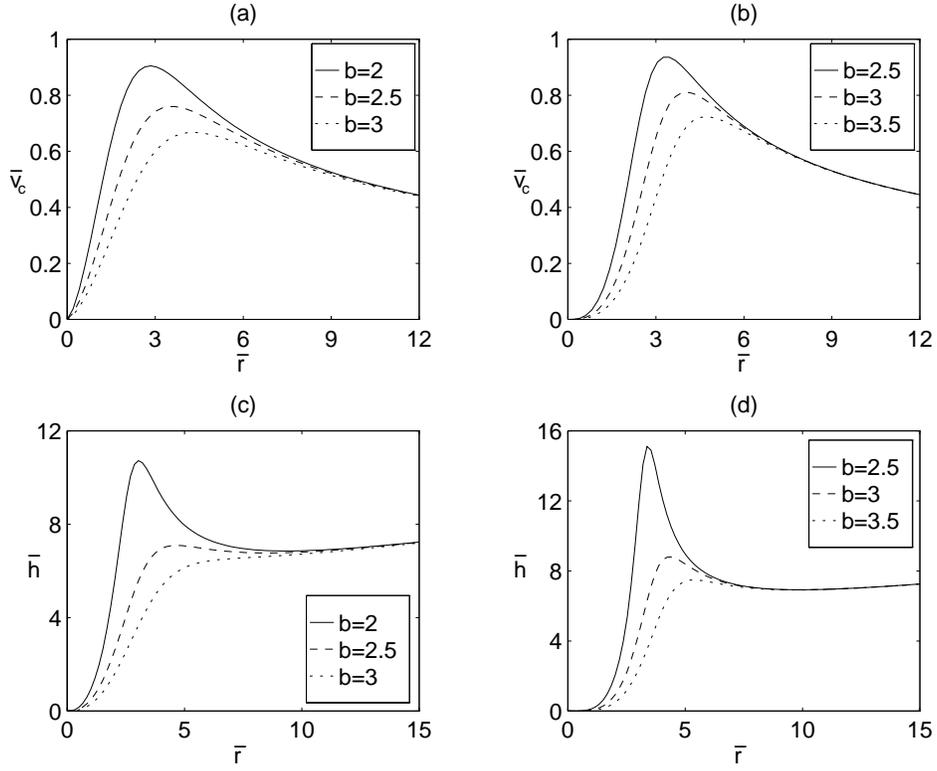}
\caption{(a)--(b) The circular velocity $\bar{v}_c=v_c/c$  (equation \ref{eq_vc_r2}),
as a function of $\bar{r}=r/r_s$ for the relativistic shell model with (a) $n=3$ and 
(b) $n=6$. (c)--(d) The angular momentum $\bar{h}=r_sc$  (equation \ref{eq_h_r2}), 
as a function of $\bar{r}=r/r_s$ for the relativistic shell model with (c) $n=3$ and
(d) $n=6$.} \label{fig4}
\end{figure}
Some curves of the circular velocity $\bar{v}_c=v_c/c$ and angular momentum 
$\bar{h}=r_sc$ for relativistic shells are shown in Figs \ref{fig4}(a)--(d). 
In Figs \ref{fig4}(a)--(b), the curves of rotation are displayed for the shells with
$n=3$ and $n=6$, respectively, and the corresponding curves of angular momentum 
are plotted in Figs \ref{fig4}(c)--(d). As the values of the parameter $\bar{b}=b/r_s$
is increased, the velocities becomes less relativistic. The minimum value 
of $\bar{b}$ such that the rotation curve is sublumial has to be found numerically.
Table \ref{tab1} shows some approximate minimum values for $\bar{b}$ for shells 
with $n=3$ to $n=9$. Note that these minimum values are always larger than 
those required by the energy conditions.
\begin{table} 
\begin{tabular}{lccccccc}
\hline
$n$ & 3 & 4 & 5 & 6 & 7 & 8 & 9 \\ 
$\bar{b}_{min.}$ & 1.78 & 2.00 & 2.18 & 2.32 & 2.43 & 2.52 & 2.60 \\
\hline 
\end{tabular}
\caption{Approximate minimum values of the parameter $\bar{b}=b/r_s$ that ensures 
sublumial rotation curves for some values of $n$.}  \label{tab1}
\end{table}

\begin{figure}
\centering
\includegraphics[scale=0.72]{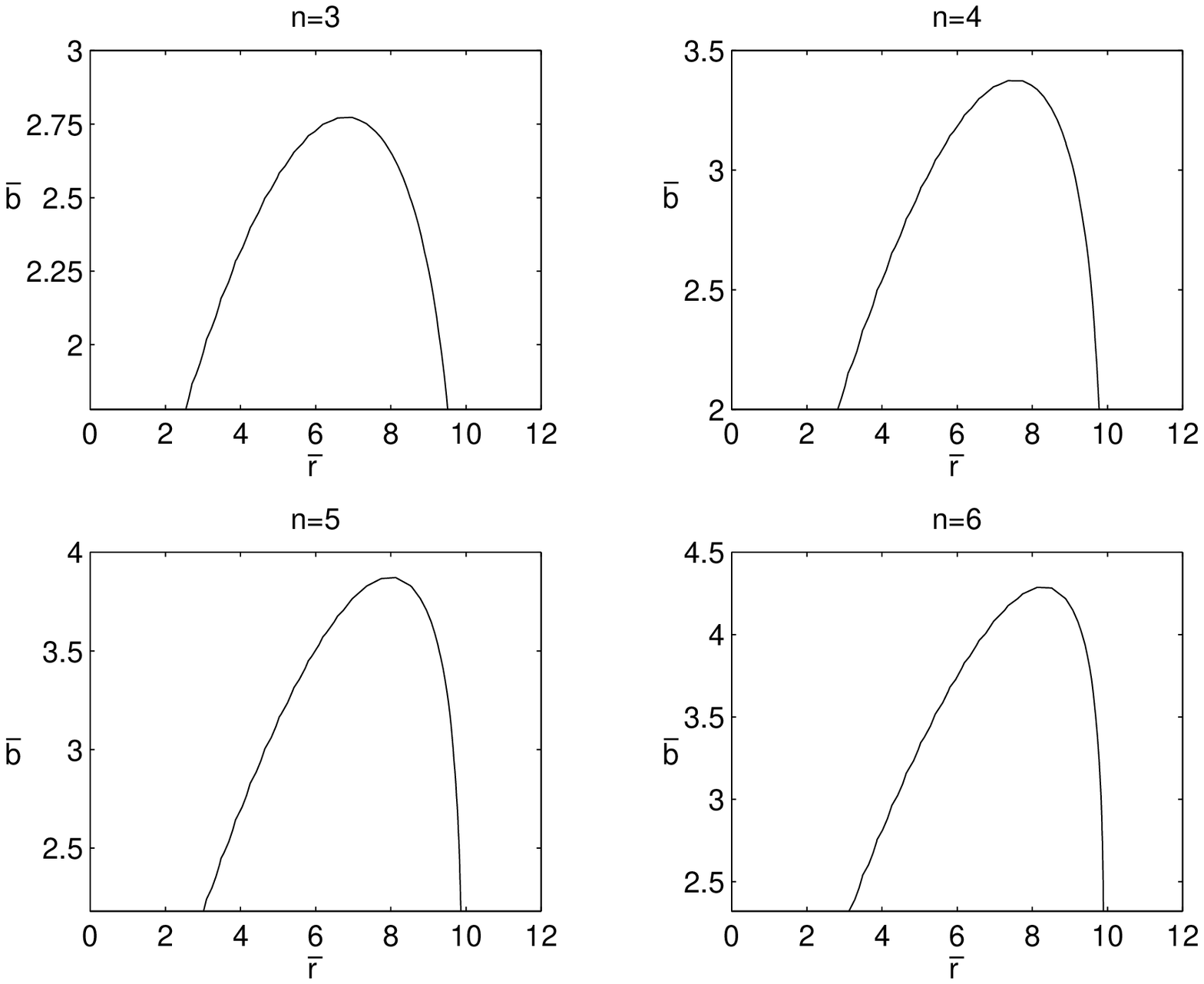}
\caption{Curves of $\mathrm{d}\bar{h}/\mathrm{d}\bar{r}=0$ as functions of
$\bar{r}$ and $\bar{b}$ for circular geodesic orbits in shells with $n=3$ to $n=6$.} \label{fig5}
\end{figure}
The curves of angular momentum in Figs \ref{fig4}(c)--(d) show regions of instability 
at high relativistic velocities. In the Newtonian
limit, the orbits are always stable, thus this instability is a purely relativistic effect. Using 
the Rayleigh criterion of stability, curves of  $\mathrm{d}\bar{h}/\mathrm{d}\bar{r}=0$
as functions of $\bar{r}$ and $\bar{b}$ are shown in Fig.\ \ref{fig5} for models of shells 
with $n$ from $n=3$ to $n=6$. In each graph, the minimum value of $\bar{b}$ is that 
given in Table \ref{tab1}. In the inner part of the curves, orbits are unstable. The 
regions of instability become larger as the parameter $n$ is increased. 
\section{Discussion} \label{sec_dis}

We proposed a very simple potential-density pair in Newtonian gravity that describes a family of spherical 
shells whose thickness is function of a parameter $n$. In the limit $n \rightarrow \infty$, one obtains 
an infinitesimal thin shell. This family of shells is found to be stable by the Rayleigh criterion of stability. 
We also studied a General Relativistic version of the Newtonian model of shells by using a particular metric in isotropic form 
in spherical coordinates. These relativistic shells have radial pressures that are different from the equal 
azimuthal and polar pressures, but are proportional to for a fixed value of $n$. Also, parameters can be
chosen such that the matter in the shells satisfies all the energy conditions. The Rayleigh criterion of stability 
shows that the relativistic shells have regions of unstable circular orbits of test particles, and this 
instability is a pure relativistic effect.   

\section*{Acknowledgments}

D.\ V.\ thanks FAPESP for financial support, and P.\ S.\ L.\ thanks FAPESP and CNPq
for financial support.

\end{document}